\documentclass[preprint,prd]{revtex4-1}
\usepackage{color}
\usepackage{graphicx}
\usepackage{epsfig}
\usepackage{subfig}

\usepackage{graphicx}% Include figure files
\usepackage{dcolumn}% Align table columns on decimal point
\usepackage{bm}% bold math

\begin{document}

\title{Regular Black Hole Solutions
 of  the Non-minimally Coupled $Y(R)F^2$ Gravity}
% Force line breaks with \\

\author{ \"{O}zcan SERT}
\email{osert@pau.edu.tr}
 \affiliation{Department of Mathematics, Faculty of Arts and Sciences,
Pamukkale
University,  20070   Denizli, T\"{u}rkiye}
%\author{Muzaffer ADAK}
 %\email{madak@pau.edu.tr}
%\affiliation{Department of Physics,
%Pamukkale
%University, 20070,  K{\i}n{\i}kl{\i},  Denizli, Turkey}


\date{\today}% It is always \today, today,

\begin{abstract}

 \noindent
   In this study we investigate regular black hole solutions of  the non-minimally coupled  $ Y(R)F^2 $  gravity model.
   We give two regular black hole solutions and the corresponding non-minimal model for both electrically  or magnetically charged cases. 
  We calculate all the energy conditions for these solutions.

%\begin{description}

% \item[PACS numbers]

%\end{description}
\end{abstract}

\pacs{Valid PACS appear here}% PACS, the Physics and Astronomy
\maketitle

%\tableofcontents

\def\ba{\begin{eqnarray}}
\def\ea{\end{eqnarray}}
\def\w{\wedge}

%\documentclass[landscape]{slides}
%\usepackage{xcolor}
%\usepackage{amsfonts}
%\usepackage{amsmath}
%\usepackage{amsbsy}
%\usepackage{amssymb,latexsym}
%\special{landscape}

%\begin{document}
%\def\ba{\begin{eqnarray}}
%\def\ea{\end{eqnarray}}
%\def\w{\wedge}

\section{Introduction}

\noindent
To understand the nature of  singularities in theories of gravity is a challenging  problem. It  can be  considered that  a  Quantum Theory of Gravity  may  solve this problem. 
As for now, we are very
far from the  Quantum Theory of Gravity, we can avoid the singularities with  regular black hole solutions.
The  regular black hole solution
 first were given   by Bardeen \cite{Bardeen}. Later, this solution of Bardeen was obtained from the field equations of the 
 Einstein-Nonlinear electrodynamics
\cite{Beato}.
It is interesting to obtain 
some  regular black hole solutions 
of 
the  theory which is $f(R)$ minimally coupled to the Non-linear electrodynamics.
In recent years, 
various new regular black holes  were proposed and investigated increasingly in the literature \cite{Hawyard}-\cite{Rodrigues}
 (see for a review \cite{Ansoldi}).
  Since the non-minimally coupled $Y(R)$-Maxwell  models
  \cite{dereli1,dereli2,Liu,bamba1,bamba2,Sert,dereli3,dereli4,Sert2}
  have some solutions which
  can explain the rotation curves of galaxies 
  and cosmic acceleration    of the universe, 
then; it is natural to seek regular black hole solutions of  the non-minimally
coupled electromagnetic fields to gravity.
 We focus on this subject in this paper.

According to the 
Penrose-Hawking singularity theorem \cite{Hawking},
 to arise 
   a singularity inside the horizon of a     black hole,
  the strong energy condition (SEC)
  has to be satisfied. The regular black holes violate   the strong energy condition  in the central region inside the black hole.
   We find various   models with the non-minimally coupled 
   $Y(R)$ function
   for some known regular metric functions. We calculate the energy conditions for the effective energy-momentum tensor of these models. Then we find that 
 they lead to a negative tangential pressure in the central core,
and 
 the effective equation of state with negative radial pressure $p_r= -\rho$ is everywhere, which is important for 
the accelerated expansion  phase of the Universe. We see that at least SEC is violated  by these solutions in some central regions of  the black holes.

\section{ The Gravitational Model with $Y(R)F^2 $-Type Coupling } \label{model}

\noindent We start with  the action with the  $Y(R)F^2 $-type non-minimal coupling term \cite{dereli3,dereli4,bamba2}
\begin{equation}
        I[e^a,{\omega^a}_b,F] = \int_M{\{   \frac{1}{2\kappa^2} R*1 -\frac{1}{2}Y(R) F\w *F  + \lambda_a\wedge T^a \} }.
        \nonumber
\end{equation}
Here  $\{e^a\}$ is the co-frame 1-form,
  ${\{\omega^a}_b\}$ is  the  connection 1-form,  $F=dA=  \frac{1}{2} F_{ab} e^a  \w  e^b $ is the homogeneous electromagnetic
  field 2-form,
$\lambda_a $ is the  Lagrange multiplier 2-form whose variation leads to the torsion-free Levi-Civita connection.
 Then the connection can be found from $T^a =  de^a + \omega^{a}_{\;\;b} \w e^b=0$.   In this action,  $R$  is the curvature scalar which can be obtained by this operation  \   $    \iota_{ba} R^{ab}= R $  from the curvature tensor 2-forms  
$ R^{a}_{\;\;b} = d\omega^{a}_{\;\;b} + \omega^{a}_{\;\;c} \w \omega^{c}_{\;\;b} $ via   the interior product $\imath_a $,  and 
 $\kappa^2 = 8\pi G$ is   universal gravitational coupling constant.   We take the  space-time metric $g = \eta_{ab} e^a \otimes
  e^b$  with the signature  $(-+++)$.
  We set  the orientated volume element as 
  $*1 = e^0 \w e^1 \w  e^2 \w e^3 $.  

  We obtain  gravitational and electromagnetic field equations of the theory    by taking  infinitesimal variations of  the action according to  independent variations of
    $\{e^a\}$,
   ${\{\omega^a}_b\}$ and $\{ A\}$ 
    \cite{dereli3,dereli4}
 \begin{eqnarray}\label{einstein}
  - \frac{1}{2 \kappa^2}  R^{bc}
 \w *e_{abc} =    && \frac{1}{2} Y  (\iota_a F \w *F - F \w \iota_a *F)   + \frac{1}{2}Y_R F_{mn} F^{mn}*R_a 
 \nonumber
 \\
 && + \frac{1}{2}  D [ \iota^b D(Y_R F_{mn} F^{mn} )]\wedge *e_{ab}
 \  ,
 \end{eqnarray}
 \begin{eqnarray}\label{maxwell1}
 d(*Y F) = 0 \; , \hskip 3 cm \  \   \   \   dF=0 \; 
 \end{eqnarray}
 where  $Y_R = \frac{dY}{dR}$. The gravitational field equation (\ref{einstein}) can be  written as 
 \begin{eqnarray}\label{Ga}
\frac{G^a}{\kappa^2} =  \tau^a 
 \end{eqnarray}
 where $G_a =- \frac{1}{2 }  R^{bc}
 \w *e_{abc} = *R_a -\frac{1}{2}R*e_a $ is the Einstein tensor, and $\tau_a = \tau_{a,b} *e^b $ is the effective energy momentum tensor for this non-minimally coupled model,  which is equal to right hand side of (\ref{einstein}). The effective energy density, radial  pressure,  and tangential  pressures are found from   $  \rho =  \tau_{0,0} $, \ 
  $  p_r =\tau_{1,1} $, \ 
   $ p_t = \tau_{ 2,2} = \tau_{3,3} $
using the field equation (\ref{Ga}).
 \section{Regular Black Hole Solutions}

 \noindent We seek regular black hole solutions for the following   (1+3)-dimensional  spherically symmetric  static line element  
 \begin{equation}\label{metric}
 g = -f^2(r)dt^2  +  f^{-2}(r)dr^2 + r^2d\theta^2 +r^2\sin(\theta)^2 d \phi^2
 \end{equation}
 and consider   the electromagnetic tensor $F$ which may have electric and  magnetic components
 \begin{eqnarray}\label{electromagnetic1}
 F   &=&  E(r) dr\wedge dt +     B(r)r^2 \sin\theta d \theta \wedge  d \phi  = E(r) e^1\wedge e^0 +   B(r)e^2 \wedge e^3.
 \end{eqnarray}
 The field equations of the model for these ansatz turn out to  five  equations (three gravitational, two electromagnetic) and four   unknown functions ($E, B, Y, f$). These equations can be found in \cite{sertd}. The homogeneous electromagnetic field condition $dF= 0$ determines the magnetic field as
 \begin{eqnarray}\label{Bq}
 B=\frac{q}{r^2}
 \end{eqnarray}
where  $q $ is a real integration constant representing  magnetic monopole charge. It is impossible  to solve these equations without any simplification. Then, we use the following constraint to simplify these equations
 \begin{eqnarray}\label{cond}
 Y_R( E^2-  B^2) =  \frac{1}{ \kappa^2}\ .
 \end{eqnarray}
 Under this constraint the number of  equations decreases to two
 \begin{eqnarray}\label{Yf}
 {f^2}'' -\frac{2}{r^2}(f^2 -1 )  = \kappa^2 Y (E^2 +B^2),
 \end{eqnarray}
 \begin{eqnarray}\label{YE}
 YE = \frac{q_e}{ r^2}, 
 \end{eqnarray}
 where $q_e$ is the electric charge.  We note that one can find the constraint  (\ref{cond})      by taking
differential of the equation (\ref{Yf}).
To show this, 
 we rewrite the equation (\ref{Yf})  using the magnetic field   $B=\frac{q}{r^2}$ from  (\ref{Bq}) and the electric field  $E=\frac{q_e}{Yr^2}$  from (\ref{YE}) and
   find

\begin{eqnarray}\label{YeniY}
r^4\left( {f^2}'' -\frac{2}{r^2}(f^2 -1 )   \right) = \kappa^2( \frac{q_e^2}{Y}  + q^2 Y)\ .
\end{eqnarray}
After taking differential  of equation (\ref{YeniY})  we obtain 
\begin{eqnarray}\label{difY}
\left(- {f^2}''' - \frac{4}{r} {f^2}'' + \frac{2}{r^2} {f^2}' + \frac{4}{r^3} (f^2 -1)\right) dr= \kappa^2(\frac{q_e^2}{Y^2r^4} -\frac{q^2}{r^4}    )dY \ .
\end{eqnarray}
We see that left hand side of the equation (\ref{difY}) is equal to differential of the curvature scalar  $R= - {f^2}'' - \frac{4}{r} {f^2}' - \frac{2}{r^2} (f^2 -1)$. Using the  equations (\ref{Bq}), (\ref{YE}) and $dR$ in (\ref{difY})  we obtain the following equation
\begin{eqnarray}
dR = \kappa^2(E^2 - B^2)dY
\end{eqnarray}
which is  differential form of the constraint (\ref{cond}).
 Thus,  it is obvious that we have two equations  (\ref{Yf}), (\ref{YE}) and three unknowns $(f, Y, E)$. Then, for a given model with a non-minimal function Y(R), we can  determine the metric function $f(r)$ from these equations. On the other hand, for a desiring metric we can reach a corresponding  model with a  non-minimal Y(R) function. In order to be  successful in this process,  we need  to solve  $r$  from $R(r)$ and  express the function Y depending   on $R$. Then we can determine the corresponding model.

\subsection{Regular Black Hole Solution-1}
 
 The   field  equations of this non-minimal $Y(R)F^2$ model  (\ref{Yf}), (\ref{YE}) accept
 the following  regular black hole   solution 
 \begin{eqnarray}\label{metric1}
 f^2(r) = 1 - \frac{2 m }{r} \left( 1 -  \frac{1}{(1+ a^3 r^3)^{1/3}    } \right)
 \end{eqnarray}
with the magnetic field ($E=0$) and the non-minimal function
\begin{eqnarray}
B(r) &=& \frac{q}{r^2}\label{B} \\
Y(r) &=& \frac{
	8 ma^6r^7 }{\kappa^2 q^2(1+ a^3r^3)^{7/3} } \label{Y1}
\end{eqnarray}
where  we have  defined a new constant  $a= \frac{2m}{q^2}$. The metric function (\ref{metric1}) can be found  in \cite{Balart}  as an electrically charged solution of Einstein-Non-linear Electrodynamics.
 In this notation, 
 we calculate  the curvature scalar $R$ and  the invariant 4-form $R_{ab}\wedge *R^{ab}$ for the metric function (\ref{metric1})
\begin{eqnarray}
R(r)  &=& \frac{8m a^3}{(1+ a^3r^3)^{7/3}} \label{R} \\
R_{ab}\wedge*R^{ab}
& =&  
(  \frac{24 m^2}{r^6} - \frac{48m^2}{r^6p}- \frac{32m^2a^6}{p^7} +
\frac{24m^2}{r^6 p^2} +
\frac{40m^2 a^6}{p^8} +
\frac{32m^2a^{12} r^6}{ p^{14}}
\nonumber \\
&& -\frac{16m^2a^3}{r^3p^4}
+\frac{16m^2a^3}{r^3p^5}  )*1
 \end{eqnarray}
where we have defined $p= (1+ a^3r^3)^{1/3}$.
When we check the limits
\begin{eqnarray}
\lim_{r\rightarrow 0 } R & =&  8m a^3 =\frac{64 m^4}{q^6}   \\
 \lim_{r\rightarrow 0 } R_{ab}\wedge*R^{ab}& =& \frac{16}{3}m^2a^6*1 = \frac{2^{10}m^8}{3q^{12}}*1,
\end{eqnarray} 
 they are regular at the center of black hole. It needs to solve $r$ from (\ref{R}) to rewrite the non-minimal function $Y$  in terms of $R$
\begin{eqnarray}\label{r}
r(R) =\frac{1}{a}\left( (  \frac{8m a^3 }{R} )^{3/7} - 1   \right)^{1/3}.
\end{eqnarray}
When we substitute the inverse function (\ref{r}) in the non-minimal function  (\ref{Y1}) we obtain
\begin{eqnarray}
Y(R)= \frac{\left[ (8ma^3)^{3/7} -R^{3/7} \right]^{7/3 }    } {\kappa^2 q^2 a^4}
\end{eqnarray}
and the
corresponding model is written as:
\begin{eqnarray}
L =     \frac{1}{2\kappa^2} R*1 - \frac{\left[ (8ma^3)^{3/7} -R^{3/7} \right]^{7/3 }    } {2\kappa^2 q^2 a^4} F\w *F  +\lambda_a\wedge T^a.
\end{eqnarray}

 After the duality transformation
 $ B\rightarrow - YE $, $q \rightarrow -q_e $ and $Y\rightarrow
 \frac{1}{Y}$;
 which is  given in  \cite{sertd},
 we   reach the field equations  of the model for  the electromagnetic tensor $F$ with only electric component ($B=0,\ E \neq  0$). As a consequence of this transformation, the same  metric function (\ref{metric1}) determines the electric field and the non-minimal function of this model as follows
 \begin{eqnarray}\label{Y11}
 Y(r) &= & \frac{\kappa^2 q_e^2 (1+ a^3r^3)^{7/3} 
 	}{8ma^6 r^7}  \\
  E(r) &=& \frac{4 q_e}{\kappa^2  r^2 } \left( 1+ \frac{1}{a^3 r^3 }\right)^{-7/3} = \frac{q_e}{Y(r)r^2}  \label{E}\; 
 \end{eqnarray}
with   $a= \frac{2m}{q_e^2}$. We see that the electric field is regular at the center of black hole, $\lim_{r\rightarrow 0 } E(r) = 0$.
 We can rewrite the non-minimal function  (\ref{Y11}) in terms of $R$ as
  \begin{eqnarray}
 Y(R)= \frac{ \kappa^2 q_e^2 a^4  } {\left[ (8ma^3)^{3/7} -R^{3/7} \right]^{7/3 }     }
 \end{eqnarray}
 and we write the corresponding Lagrangian of the model
 \begin{eqnarray}
L =     \frac{1}{2\kappa^2} R*1 - \frac{\kappa^2 q_e^2 a^4 }{2 \left[ (8ma^3)^{3/7} -R^{3/7} \right]^{7/3 }    }  F\w *F   +\lambda_a\wedge T^a
 \end{eqnarray}
via the duality transformation.

 The same metric function (\ref{metric1})   and an electric field different from (\ref{E}) only up to a scale factor was obtained from   a different theory with Einstein-nonlinear electrodynamics  in \cite{Balart}.

 It is important to check all the energy conditions for this solution.    The conditions which are calculated as below  have to be  equal or greater than zero for rising  a singularity  in  General Relativity.  But in the regular black holes at least the strong energy condition has to be violated. To show this
  we firstly calculate the energy density $\rho(r) $, the radial and tangential pressures $p_r, p_t$ for the metric function ($\ref{metric1}$) and the magnetic field (\ref{B}). We note that all the following results also can be obtained  from the solutions with  the electric field (\ref{E}) which has the electric charge $q_e=q$

  \begin{eqnarray}
  \rho(r) =  \frac{ 16 m^4 q^2 }{\kappa^2 ( q^6 + 8 m^3r^3)^{4/3}  }
  =
  -p_r(r) 
  \hskip 2 cm
  p_t(r) = \frac{16 q^2m^4( 8 m^3 r^3 - q^6  ) }{ \kappa^2 ( q^6 + 8 m^3r^3)^{7/3}  } \; .
  \end{eqnarray}
  We find the following energy conditions   using the energy density $\rho(r)$, the radial and tangential  pressures  $p_r, p_t$
  \begin{eqnarray}
  DEC_1 & = & \rho   \geq 0,
  \\
  NEC_1 &  = & WEC_1 = \rho + p_{r} = 0\ ,
  \\
  NEC_2 &  = &  WEC_2 = \rho + p_{t} = 
  \frac{ 2^8 m^7q^2 r^3}{
  	\kappa^2 ( q^6 + 8 m^3r^3)^{7/3}
  	} \ , 
  \\
  SEC \ \ & = &  \rho + p_r + 2p_t =  \frac{32m^4q^2(8m^3 r^3 -q^6)}{ 	\kappa^2 ( q^6 + 8 m^3r^3)^{7/3} }  \ ,
  \\
  DEC_2 & = & \rho -  p_r = 2\rho \ ,
  \\
  DEC_3 & = & \rho -  p_t = \frac{32 m^4 q^8 }{\kappa^2 ( q^6 + 8 m^3r^3)^{7/3}} \ .
  \end{eqnarray}
 Thus, we see that all the energy conditions are satisfied in the region $r\geq \frac{q^2}{2m}$ for the electrically or magnetically charged solutions.
 But, only the SEC is violated in the central  region $r < \frac{q^2}{2m} $.

\subsection{Regular Black Hole Solution-2}
 
 Secondly, we take the metric function from  \cite{Balart2} and   \cite{Rodrigues}  with the electric charge $q_e$
 \begin{eqnarray}\label{m2}
 f^2(r) = 1- \frac{2m}{r} e^{-\frac{q_e^2}{2mr}}\ .
 \end{eqnarray}
 
The Ricci scalar and $R_{ab}\wedge *R^{ab} $    are calculated using the metric function (\ref{m2}) as 
 \begin{eqnarray}\label{R2}
 R(r) =  \frac{q_e^4}{2mr^5} e^{-\frac{q_e^2}{2mr}} 
 \end{eqnarray}
 \begin{eqnarray} 
R_{ab}\wedge*R^{ab} =  (  \frac{24 m^2}{r^6} - \frac{24m q_e^2}{r^7} + \frac{12 q_e^4 }{r^8} - 
 \frac{2q_e^6}{mr^9 } +
 \frac{ q_e^8}{8m^2 r^{10} }  )e^{-\frac{q_e^2}{mr}}*1 \ .
 \end{eqnarray}
 
 When we check their limits 
 \begin{eqnarray}
 \lim_{r\rightarrow 0 } R = 0, \hskip 1 cm \lim_{r\rightarrow 0 } R_{ab}\wedge*R^{ab} =0  
 \end{eqnarray}
we see that  they are regular at the center of black hole.
 We find the solution of these differential equations (\ref{Yf}) and (\ref{YE}) for this regular metric function (\ref{m2}) as follows
 \begin{eqnarray}
 Y(r)& = & \frac{  2\kappa^2 mr}{ (8mr -q_e^2)e^{-\frac{q_e^2}{2mr}}  }\label{Y2} \\
 E(r) & = &\frac{q_e}{r^2Y(r)} =  \frac{ q_e(  8m r - q_e^2 ) e^{-\frac{q_e^2}{2mr }}   }{ 2m\kappa^2 r^3} \label{E2}\ .
 \end{eqnarray}
 While this electric field is regular at the center,  it has  the following  asymptotic  behavior   
  \begin{eqnarray}
 E(r) =  \frac{4q_e}{\kappa^2 r^2} - \frac{5q_e^3}{2 m\kappa^2r^3} +\frac{3q_e^5}{4m^2\kappa^2 r^4} + O(\frac{1}{r^5})\ .
 \end{eqnarray}
  
  The inverse function of  $R(r)$ in (\ref{R2}) can be found  in terms of Lambert function \cite{Dence} as
  \begin{eqnarray}
  r(R) = -\frac{q_e^2}{10 m  } W^{-1} \left[- \frac{1}{10} \left(  \frac{2 q_e^6R}{ m^4} \right)^{1/5} \right]\ .
  \end{eqnarray}
  Then we rewrite the non-minimal function of this model as
   \begin{eqnarray}
   Y(R) = - \frac{}{}\frac{10^5 \kappa^2 m^4 W^5 \left[ -\frac{1}{10} \left(  \frac{2q_e^6 R}{m^4}\right)^{1/5}\right]}{ 2q_e^6 R \left( 4+ 5W\left[ - \frac{1}{10} \left(  \frac{2q_e^6 R}{m^4}\right)^{1/5}\right] \right) }\ .
   \end{eqnarray}
   From  the duality transformation $YE\rightarrow  - B $,\  \ $q_e\rightarrow -q$, \  $Y \rightarrow \frac{1}{Y}$ \ \cite{sertd}, we can find the  magnetic solution for this same metric (\ref{m2}), which corresponds to dual solution of the electrically charged solution (\ref{Y2}), (\ref{E2}). Then
the resulting magnetic field
  \begin{eqnarray}\label{B2}
  B(r) = \frac{q}{r^2}
  \end{eqnarray}
  and the corresponding non-minimal function
    \begin{eqnarray}
    Y(R) = -    \frac{ 2q^6 R \left( 4+ 5W\left[ - \frac{1}{10} \left(  \frac{2q^6 R}{m^4}\right)^{1/5}\right]\right)   }{ 10^5 \kappa^2 m^4 W^5 \left[ - \frac{1}{10} \left(  \frac{2q^6 R}{m^4}\right)^{1/5}\right] }
    \end{eqnarray}
    constructs the dual solution.
  We calculate the energy density and pressures for the metric function (\ref{m2}) with  the magnetic charge $q$ and the  the magnetic field  (\ref{B2}).
   We note that all the following results also can be obtained  from the solutions with  the electric field (\ref{E2}) which has the electric charge $q_e=q$
   \begin{eqnarray}
   \rho(r) =  \frac{q^2 e^{\frac{-q^2}{2 mr }}}{\kappa^2 r^4}
   =
      -p_r(r) 
       \hskip 2 cm
       p_t(r) = \rho(r)  -   \frac{q^4 e^{\frac{-q^2}{2 mr }}}{4 \kappa^2 m r^5} \ .
   \end{eqnarray}

Now we calculate  all the energy conditions for the metric function (\ref{m2}) 
\begin{eqnarray}
DEC_1 & = & \rho   \geq 0,
\\
NEC_1 &  = & WEC_1 = \rho + p_{r} = 0\ ,
 \\
NEC_2 &  = &  WEC_2 = \rho + p_{t} = 
\frac{q^2 e^{\frac{-q^2}{2 mr }}}{4 m \kappa^2 r^5} (8mr -q^2  ) \ , 
\\
SEC \ \ & = &  \rho + p_r + 2p_t =  \frac{q^2 e^{\frac{-q^2}{2 mr }}}{2 m \kappa^2 r^5} (4mr -q^2  )  \ ,
\\
DEC_2 & = & \rho -  p_r = 2\rho \ ,
\\
DEC_3 & = & \rho -  p_t =  \frac{q^4 e^{\frac{-q^2}{2 mr }}}{4 \kappa^2 m r^5} \ .
\end{eqnarray}
 
  We found that all the energy conditions are satisfied in the region $r \geq \frac{q^2}{4 m}$ for these electrically ($q_e=q$) or magnetically  charged solutions. But,   in the region $ \frac{q^2}{8m } \leq r  <  \frac{q^2}{4m }$ only the SEC is violated by this solution. Furthermore,
  in the central region $r < \frac{q^2}{8m }$  the conditions  $NEC_2, WEC_2$ together with  SEC  are violated.

  %%%%%%%%%%%%%%%%%

\section{Conclusion}

\noindent
 We   have investigated  various  regular black hole solutions of the non-minimally coupled $Y(R)F^2$
 theory.  We found  electrically charged   or magnetically charged (after the duality transformation)   regular black hole solutions  which can be obtained  from the non-minimal model with some specific non-minimal functions $Y(R)$.
 We calculated all the energy conditions  for these solutions
 using the effective energy-momentum tensor that comes from  the non-minimally coupled $Y(R)F^2$ term.
 
The first regular black hole solution violates only the strong energy condition in a central region  $r < \frac{q^2}{2m} $,  inside the event horizon. 
This solution is in agreement  with 
the singularity theorem of General Relativity \cite{Zaslavski}.
But  the second regular black hole solution violates the weak energy condition  together with   the strong energy condition  in  the region $r < \frac{q^2}{8m} $, while it satisfies  all the energy  conditions in the outer  region  $r \geq \frac{q^2}{4m} $ for the electric or magnetic fields.
The same energy conditions  of the second regular black hole  with an electric field are also  found in \cite{Rodrigues} for a different theory 
which is $f(R)$ minimally coupled to the Non-linear electrodynamics.

%It is interesting to find  other regular black hole solutions and  to investigate the properties of them, for the next studies.

%\newpage

\end{document}